\documentclass[3p,times,procedia]{elsarticle}
\usepackage{nupha_ecrc}
\usepackage{graphicx}
\usepackage{caption}
\usepackage{subcaption}
\biboptions{numbers,sort&compress}

\volume{00}

\firstpage{1}

\journalname{Nuclear Physics A}

\runauth{D. Moreira de Godoy}

\providecommand{\pT}{$p_{\rm T}$ }

\providecommand{\raa}{$R_{\rm{\rm AA}}$ }

\providecommand{\vtwo}{$v_{\rm 2}$ }

\providecommand{\s}{$\sqrt{s}$ }
\providecommand{\snn}{$\sqrt{s_{\rm NN}}$ }

\providecommand{\vtwoe}{$v_{\rm 2}^{ e^{\pm} \leftarrow HF}$}
\providecommand{\vtwom}{$v_{\rm 2}^{ \mu \leftarrow HF}$}

\providecommand{\raae}{$R_{\rm{\rm AA}}^{ e^{\pm} \leftarrow HF}$}
\providecommand{\raam}{$R_{\rm{\rm AA}}^{ \mu \leftarrow HF}$}

\jid{nupha}

\jnltitlelogo{Nuclear Physics A}

\usepackage{amssymb}
\usepackage{float}
\usepackage{lineno}

\usepackage[figuresright]{rotating}

\begin{document}

\begin{frontmatter}



\dochead{}

\title{Measurements of the nuclear modification factor and  elliptic flow of leptons from heavy-flavour hadron decays in Pb--Pb collisions at $\sqrt{s_{\rm NN}}$ = 2.76 and 5.02 TeV with ALICE}


\author{Denise Moreira de Godoy on behalf of the ALICE Collaboration}
\address{Westf\"alische Wilhelms-Universit\"at M\"unster, Wilhelm-Klemm-Str. 9, M\"unster, Germany}

\begin{abstract}
We  present the ALICE results on the nuclear modification factor and   elliptic flow of electrons and muons from open heavy-flavour hadron decays at mid-rapidity and forward rapidity in Pb--Pb collisions at $\sqrt{s_{\rm NN}}$ = 2.76 and 5.02 TeV for different centrality intervals. 
The results are compared to model calculations that include interactions of heavy quarks with the medium.
\end{abstract}



\end{frontmatter}




\section{Introduction}
\label{Sec:Introduction}

Heavy quarks, i.e. charm and beauty, are sensitive probes to study the properties of the strongly-interacting matter created in heavy-ion collisions at ultra-relativistic energies, since they are mainly produced in initial hard scattering processes and experience the entire evolution of the system.

The heavy quarks  traversing the medium lose energy via collisional and radiative processes in the interaction with the medium constituents. The in-medium energy loss of heavy quarks can be investigated with the nuclear modification factor ($R_{\rm AA}$) of heavy-flavour particles, which is defined as the ratio of the transverse momentum ($p_{\rm T}$) differential yield of particles in heavy-ion collisions ($\mathrm{d}N_{\rm{AA}} / \mathrm{d}p_{\rm T}$) with respect to the $p_{\rm T}$-differential cross section in pp collisions ($\mathrm{d}\sigma_{\rm{pp}}/\mathrm{d}p_{\rm T}$) scaled with the average nuclear overlap function in heavy-ion collisions ($\left\langle T_{\rm{AA}} \right\rangle$), i.e. $R_{\rm{AA}} = (\mathrm{d}N_{\rm{AA}} / \mathrm{d}p_{\rm T}) / ( \left\langle T_{\rm{AA}} \right\rangle   \mathrm{d}\sigma_{\rm{pp}}/\mathrm{d}p_{\rm T})$.
In absence of nuclear effects, $R_{\rm AA}$ is expected to be unity. The energy loss of heavy flavours in the medium causes 
a shift of the momentum distribution towards lower values, resulting in a suppression of heavy-flavour particle yields, $R_{\rm AA}$ $<$ 1, at intermediate and high $p_{\rm T}$.

Further properties of the medium can be investigated with the elliptic flow of heavy-flavour particles, $v_{\rm 2} = \langle \cos[ 2( \varphi - \Psi_{\rm 2})]\rangle$, which is defined as the coefficient of the second-order harmonic of the Fourier expansion of the distribution of the particle azimuthal angle ($\varphi$) in momentum space with respect to the angle of the 2$^{nd}$-order symmetry plane ($\Psi_{\rm 2}$) \cite{PhysRevC.58.1671}. The measurement of the heavy-flavour particle $v_{\rm 2}$ at low $p_{\rm T}$ provides insight into the collective motion of heavy quarks in the medium, while the heavy-flavour particle $v_{\rm 2}$ at high $p_{\rm T}$ is sensitive to the path-length dependence of the energy loss of heavy quarks in the almond-shaped overlap area in non-central collisions. 

The semi-leptonic decays of heavy-flavour hadrons are well suited for heavy-flavour studies, since the ALICE detector has an unique capability for identification of electrons and muons  over a wide $p_{\rm T}$ range. In addition, the contributions of charm and beauty-hadron decays can be disentangled in the yield of electrons.

\section{Analysis}
\label{Sec:Analysis}

Muons are reconstructed in ALICE  with the Muon Spectrometer \cite{alice} at forward rapidity (2.5 $< y <$ 4). 
The muons from background sources, mainly decays of $\pi$ and $\mathrm{K}$ at low-intermediate $p_{\rm T}$ and decays of $\mathrm{W}$ at high $p_{\rm T}$, are statistically subtracted from the measured muon sample. The contribution of muons from  $\pi$ and $\mathrm{K}$ decays is estimated with a data-tuned Monte-Carlo (MC) cocktail, while the contribution of muons from $\mathrm{W}$ decays is estimated with a MC simulation based on POWHEG \cite{POWHEG}. 
The \raa of muons from heavy-flavour hadron decays (\raam) at forward rapidity  has been measured in Pb--Pb collisions at \snn = 2.76 \cite{ALICEmuRPbPb} and 5.02 TeV for various centrality intervals.
The data sample used for the analysis in Pb--Pb collisions at \snn = 5.02 TeV in the 0--90\% centrality class consist of 4.7$\cdot$10$^{7}$ and 10$^{8}$ muon-triggered collisions for low- and high-\pT trigger thresholds, respectively.
The pp reference for the \raa  analysis  in  Pb--Pb collisions at \snn = 5.02 TeV  is obtained by  a $\sqrt{s}$-scaling \cite{scaling} of the measured cross section of muons from heavy-flavour hadron decays in pp collisions at  \s = 7 TeV \cite{Abelev2012265} for $p_{\rm T} <$ 12 GeV/c and by an extrapolation of the measured cross section for higher $p_{\rm T}$. 
The \vtwo of muons from heavy-flavour hadron decays (\vtwom) at forward rapidity  has been measured with the two-particle $Q$ cumulant method  in Pb--Pb collisions at \snn = 2.76 TeV  \cite{Adam:2015pga} in the 0--10\%, 10--20\%, and 20--40\% centrality classes.

Electrons are identified at mid-rapidity with the Inner Tracking System (ITS), the Time Projection Chamber (TPC), the Time Of Flight (TOF) and the ElectroMagnetic Calorimeter (EMCal) \cite{alice}. 
The contribution of electrons that do not originate from heavy-flavour hadron decays, which are mainly electrons from photon conversions in detector material and from Dalitz decays of neutral mesons, is obtained exploiting the invariant mass of electron-positron pairs and/or the cocktail method, depending on the analysis. The electron background is then statistically subtracted from the measured electron sample.
The \raa of electrons from heavy-flavour hadron decays (\raae) at mid-rapidity  ($|y| <$ 0.6) has been measured in Pb--Pb collisions at \snn = 2.76  \cite{HFERAA} and 5.02 TeV for several centrality classes. 
 The data sample used for the analysis in  Pb--Pb collisions at \snn = 5.02 TeV consist of 10$^{7}$ semi-central (30--50\%) collisions recorded with a minimum-bias trigger.
The pp reference for the \raa  analysis in  Pb--Pb collisions at \snn = 5.02 TeV is obtained by interpolating the cross sections of  electrons from heavy-flavour hadron decays  in pp collisions at  \s = 2.76 and 7 TeV, as  discussed in~\cite{Adam201681}.
The \vtwo of electrons from heavy-flavour hadron decays (\vtwoe) at mid-rapidity ($|y| <$ 0.7) has been measured with the event plane method \cite{PhysRevC.58.1671} in three centrality classes (0--10\%, 10--20\%, and 20--40\%)  in Pb--Pb collisions at \snn = 2.76 TeV \cite{Adam:2016ssk} and in  30--50\%  Pb--Pb collisions at \snn = 5.02 TeV. The V0 detector, covering the backward rapidity (V0A, 2.8 $< \eta <$ 5.1) and forward rapidity  (V0C, $-$3.7 $< \eta <$ $-$1.7) regions, is used to obtain the collision centrality and the symmetry-plane angle ($\Psi_{\rm 2}$) needed in the \vtwo analysis with the event plane method.

\section{Results}
\label{Sec:Results}

\begin{figure}[!ht]
    \centering
    \begin{subfigure}[b]{0.495\textwidth}
        \includegraphics[scale=0.107, height=5cm]{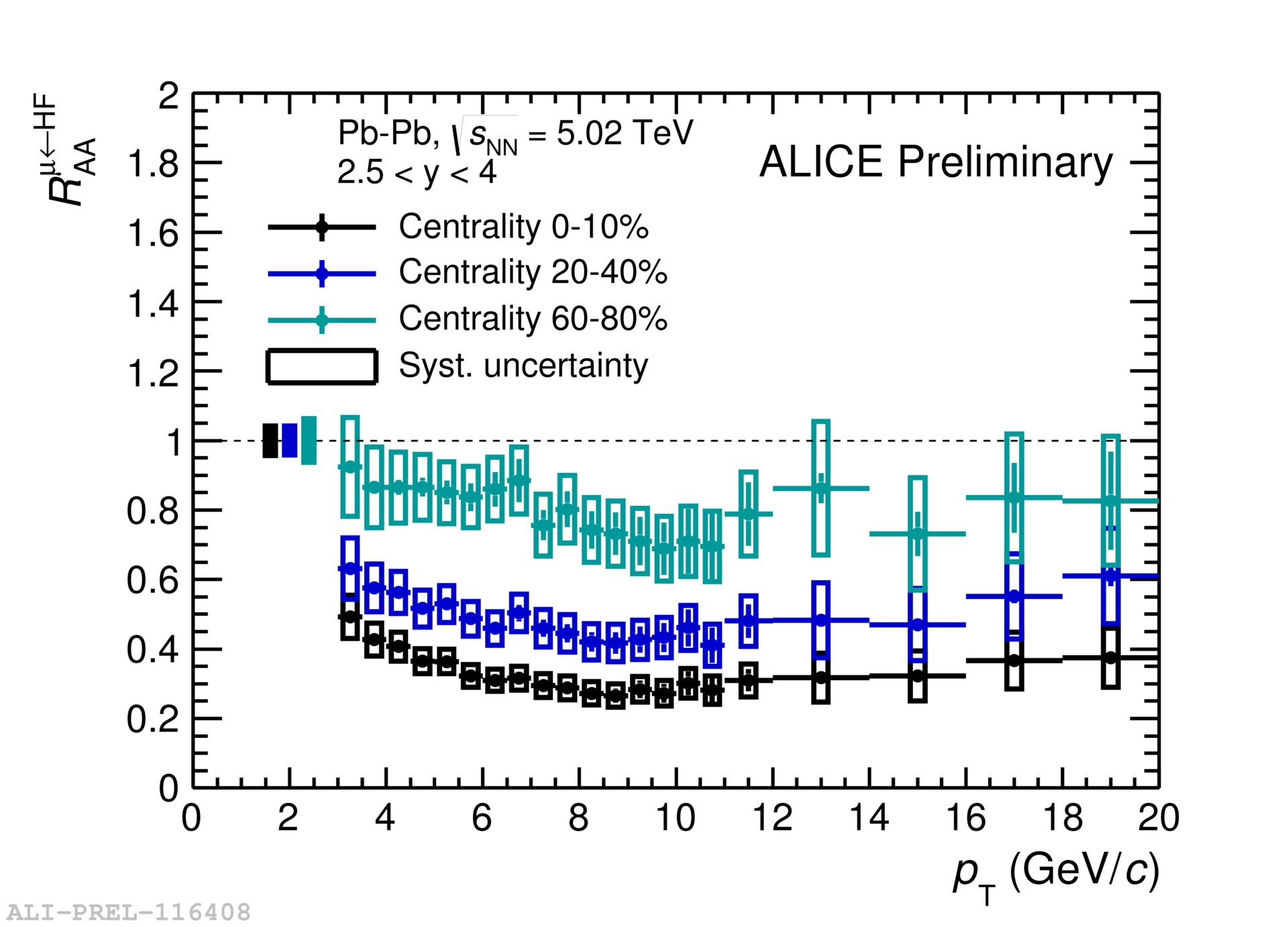}
    \end{subfigure}
        \begin{subfigure}[b]{0.495\textwidth}
        \includegraphics[scale=0.107, height=5cm]{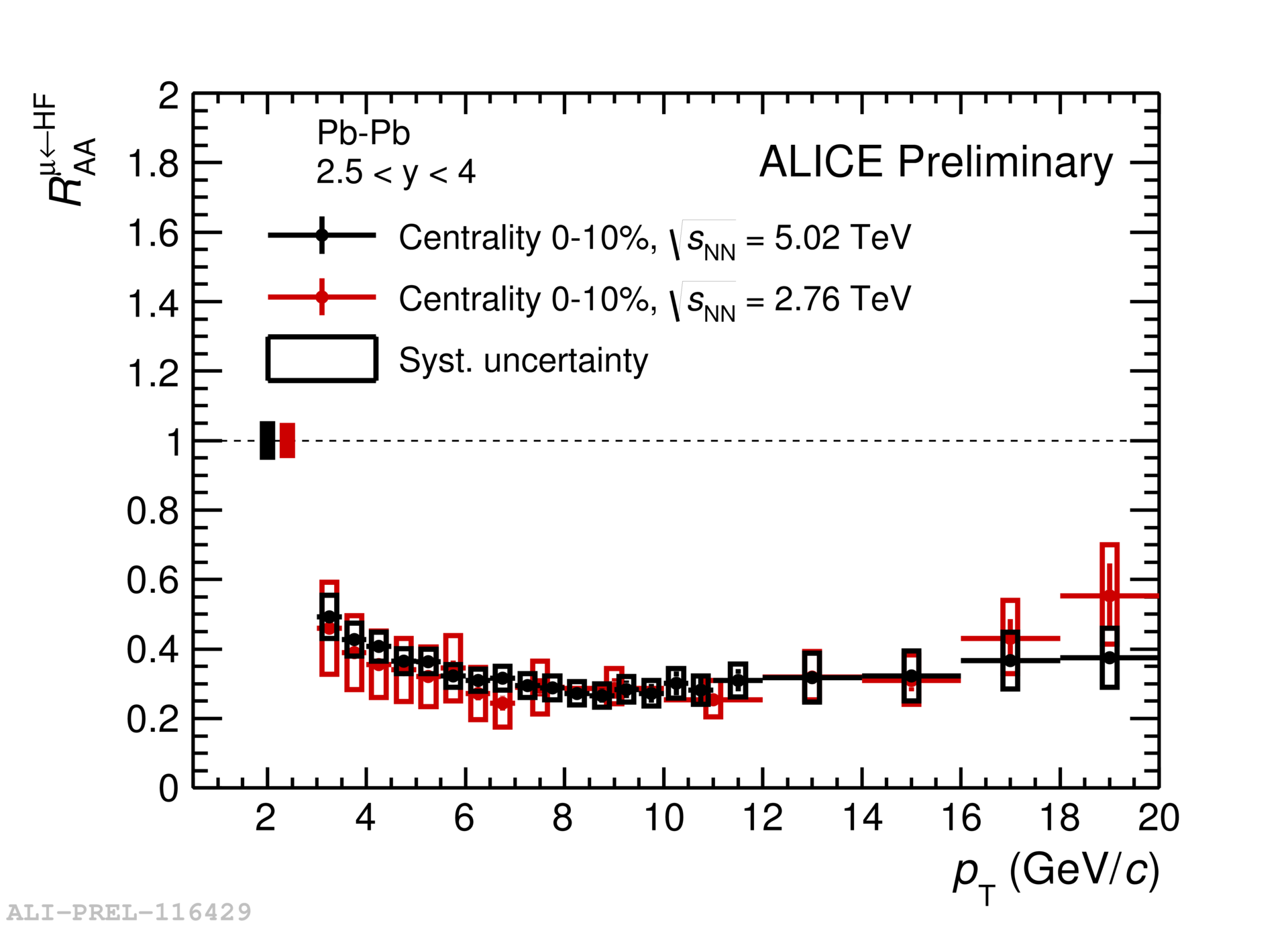}
    \end{subfigure}
    \caption{Left: \raa of muons from heavy-flavour hadron decays as a function of \pT in Pb--Pb collisions at \snn = 5.02 TeV in the 0--10\%,  20--40\% and  60--80\% centrality classes. Right:  Comparison of the \raa of muons from heavy-flavour hadron decays as a function of \pT in the 10\% most central Pb--Pb collisions at \snn = 2.76 \cite{ALICEmuRPbPb} and 5.02 TeV.}
\label{Fig:RAA_HFmrun2}
\end{figure}

\begin{figure}[!ht]
    \centering
        \begin{subfigure}[b]{0.495\textwidth}
        \includegraphics[scale=0.107, height=4.9cm]{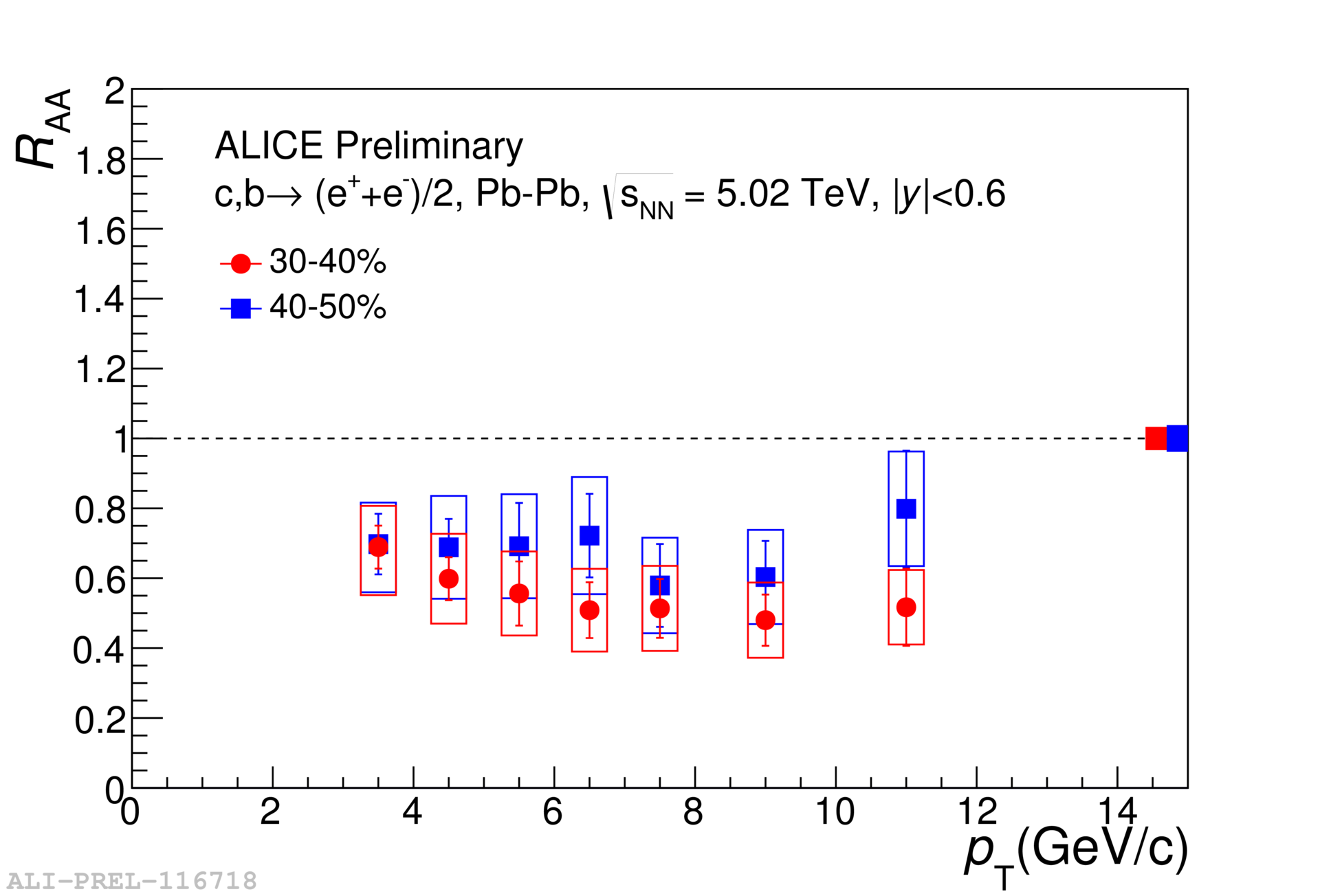}
    \end{subfigure}
    \begin{subfigure}[b]{0.495\textwidth}
        \includegraphics[scale=0.107, height=4.9cm]{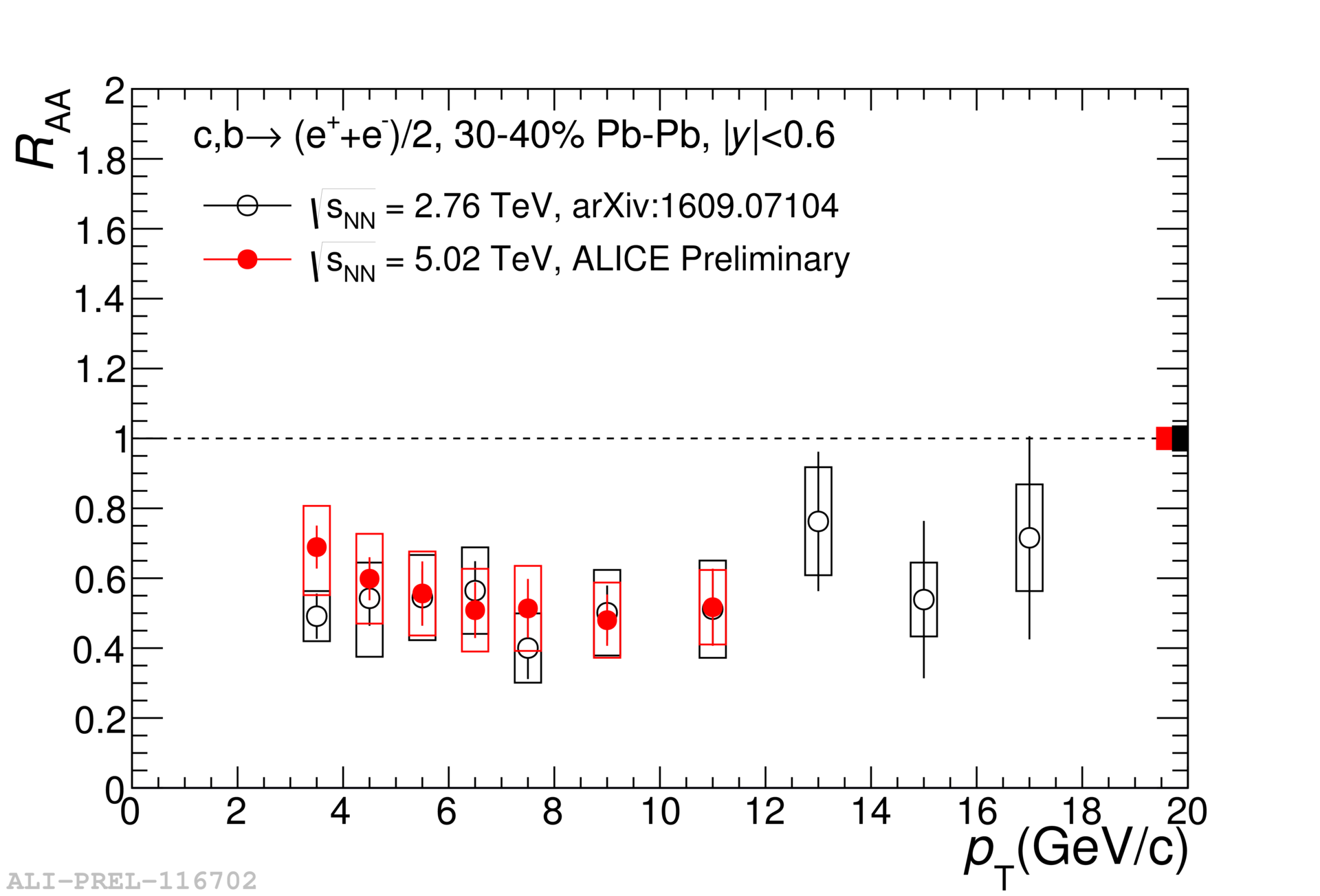}
    \end{subfigure}
    \caption{Left: \raa of electrons from heavy-flavour hadron decays as a function of \pT in Pb--Pb collisions at \snn = 5.02 TeV in the 30--40\% and 40--50\% centrality classes. Right:  Comparison of the \raa of electrons from heavy-flavour hadron decays as a function of \pT in 30--40\% Pb--Pb collisions at \snn = 2.76  \cite{HFERAA} and 5.02 TeV.}
\label{Fig:RAA_HFErun2}
\end{figure}

The  \raam at forward rapidity (2.5 $< y <$ 4) and \raae at  mid-rapidity ($|y| <$ 0.6) in Pb--Pb collisions at \snn = 5.02 TeV, as a function of \pT and centrality class, are shown in the left panels of Figures  \ref{Fig:RAA_HFmrun2} and \ref{Fig:RAA_HFErun2}, respectively.
A suppression of leptons from heavy-flavour hadron decays is observed in Pb--Pb collisions at \snn = 5.02 TeV, which is mainly induced by final-state effects due to heavy-quark energy loss in the medium since no significant modification of the spectra of leptons from heavy-flavour hadron decays is  observed in p--Pb collisions at \snn = 5.02 TeV  relative to binary scaled  pp collisions \cite{Acharya:2017hdv,Adam201681}. 
The suppression decreases from central to peripheral collisions, as expected from the centrality dependence of the size and initial energy density of the medium.
The same features are observed for Pb--Pb collisions at \snn = 2.76 TeV \cite{ALICEmuRPbPb,HFERAA}. In fact, the \raa of leptons from heavy-flavour hadron decays is compatible, within uncertainties, in Pb--Pb collisions at \snn = 2.76 and 5.02 TeV, as shown in the right panels of Figures \ref{Fig:RAA_HFmrun2} and \ref{Fig:RAA_HFErun2}. 
The \raae  at mid-rapidity ($|y| <$ 0.6) and \raam at forward rapidity (2.5 $< y <$ 4) in 0--10\% Pb--Pb collisions at \snn = 2.76 TeV are compatible within uncertainties \cite{HFERAA}.

\begin{figure}[!ht]
    \centering
    \includegraphics[scale=0.115, height=5cm]{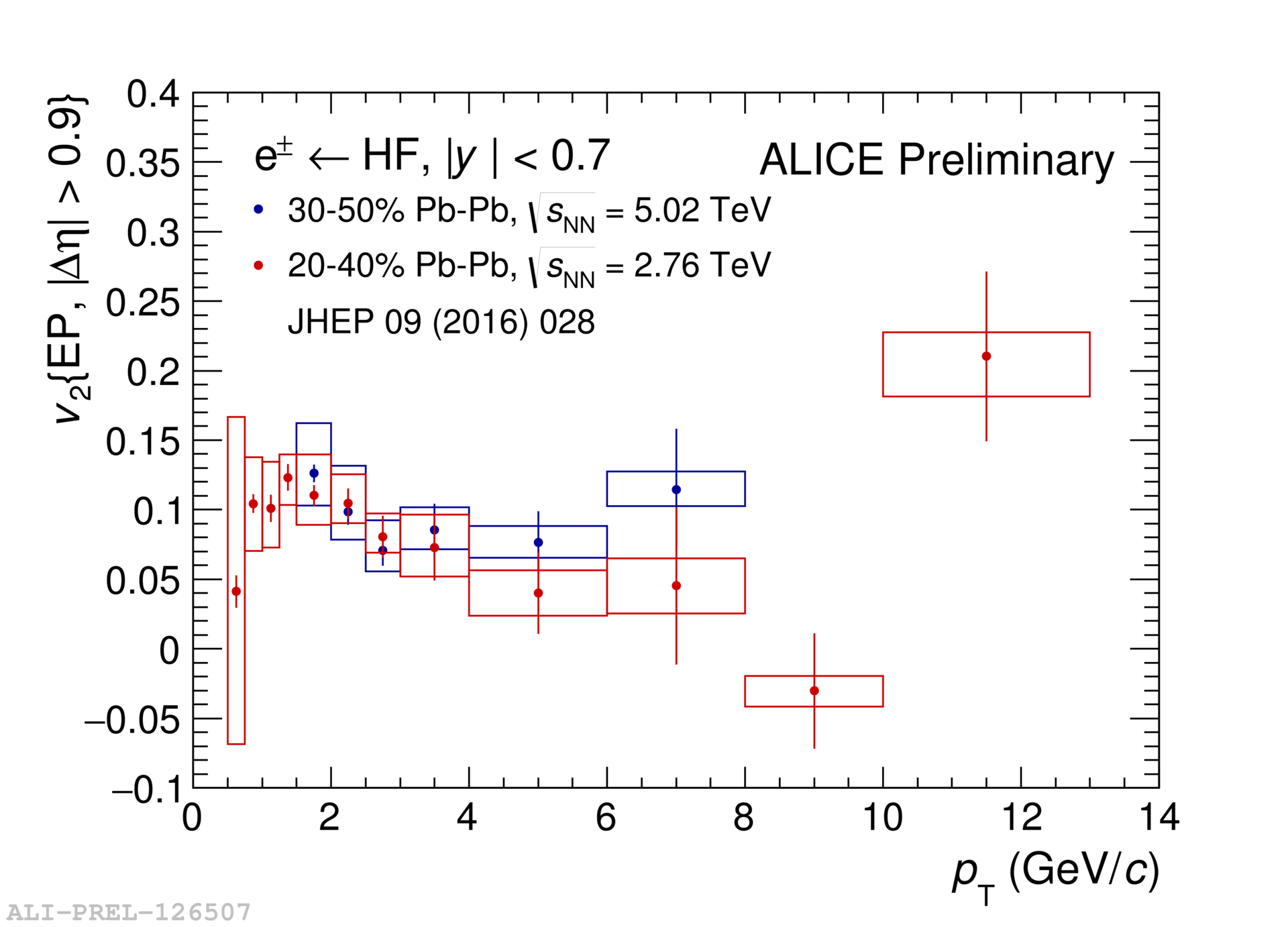}
    \caption{\vtwo of electrons from heavy-flavour hadron decays as a function of \pT in 20--40\% Pb--Pb collisions at \snn = 2.76 TeV \cite{Adam:2016ssk} and in 30--50\% Pb--Pb collisions at \snn = 5.02 TeV.}
\label{Fig:v2_HFE}
\end{figure}

The \pT dependence of the \vtwoe  exhibits the same trend  in 20--40\% Pb--Pb collisions at \snn = 2.76 TeV \cite{Adam:2016ssk} and in 30--50\% Pb--Pb collisions at \snn = 5.02 TeV, as shown in Fig. \ref{Fig:v2_HFE}.
 A positive \vtwo at low-intermediate \pT is observed in both collision energies:  at \snn = 2.76 TeV the significance has a maximum of 5.9$\sigma$ in the 2--2.5 GeV/c interval 
 and at  \snn = 5.02 TeV the maximum significance is 5.3$\sigma$ in the 1.5--2 GeV/c interval. 
  The data suggest that heavy quarks are affected by the collective motion of the system.
The \vtwoe  at mid-rapidity ($|y| <$ 0.7) and the \vtwom at forward rapidity (2.5 $< y <$ 4) in Pb--Pb collisions at \snn = 2.76 TeV are compatible within uncertainties \cite{Adam:2016ssk}.

\begin{figure}[!ht]
    \centering
    \begin{subfigure}[b]{0.49\textwidth}
        \includegraphics[scale=0.08, height=5cm]{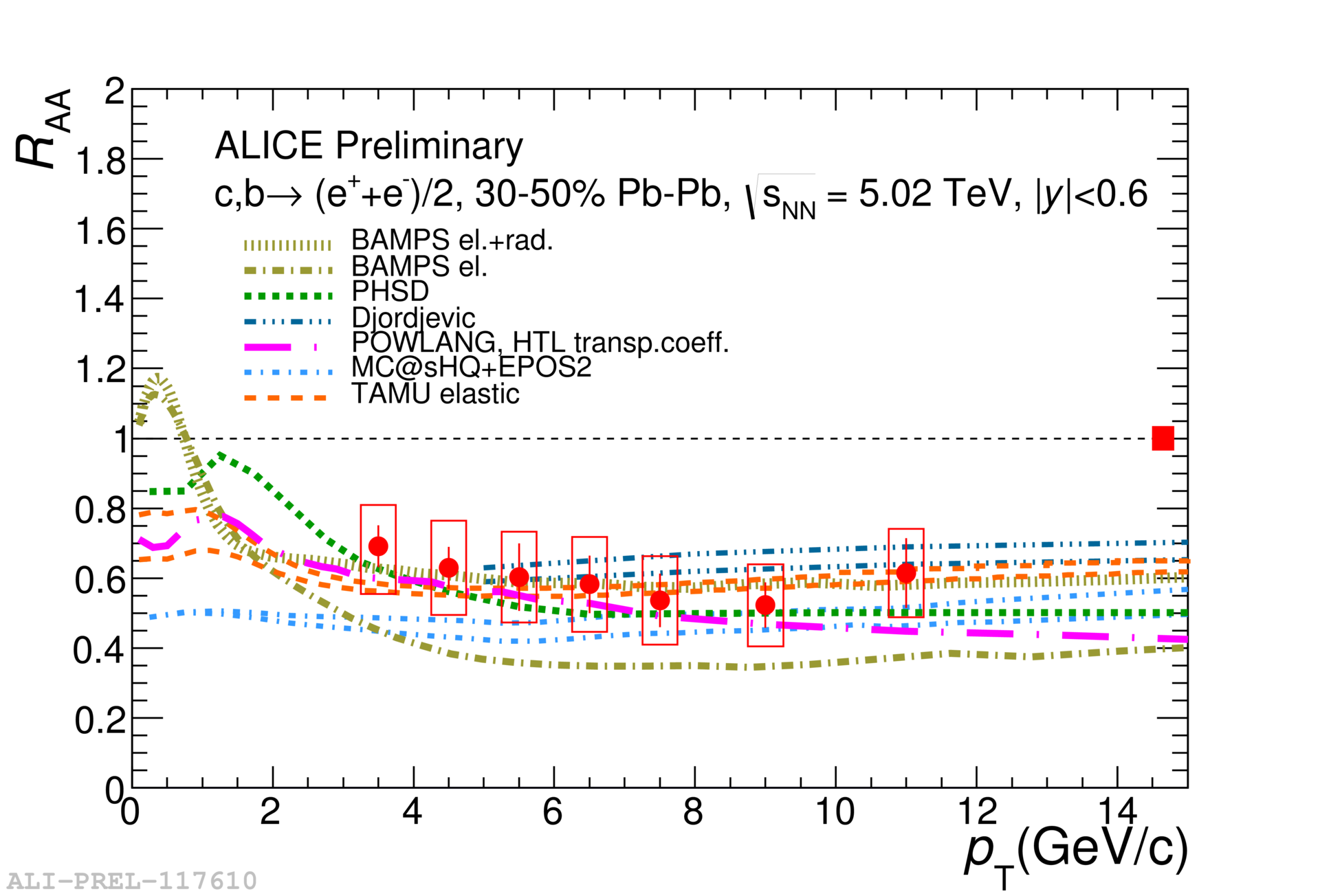}
    \end{subfigure}
        \begin{subfigure}[b]{0.49\textwidth}
        \includegraphics[scale=0.13, height=5cm]{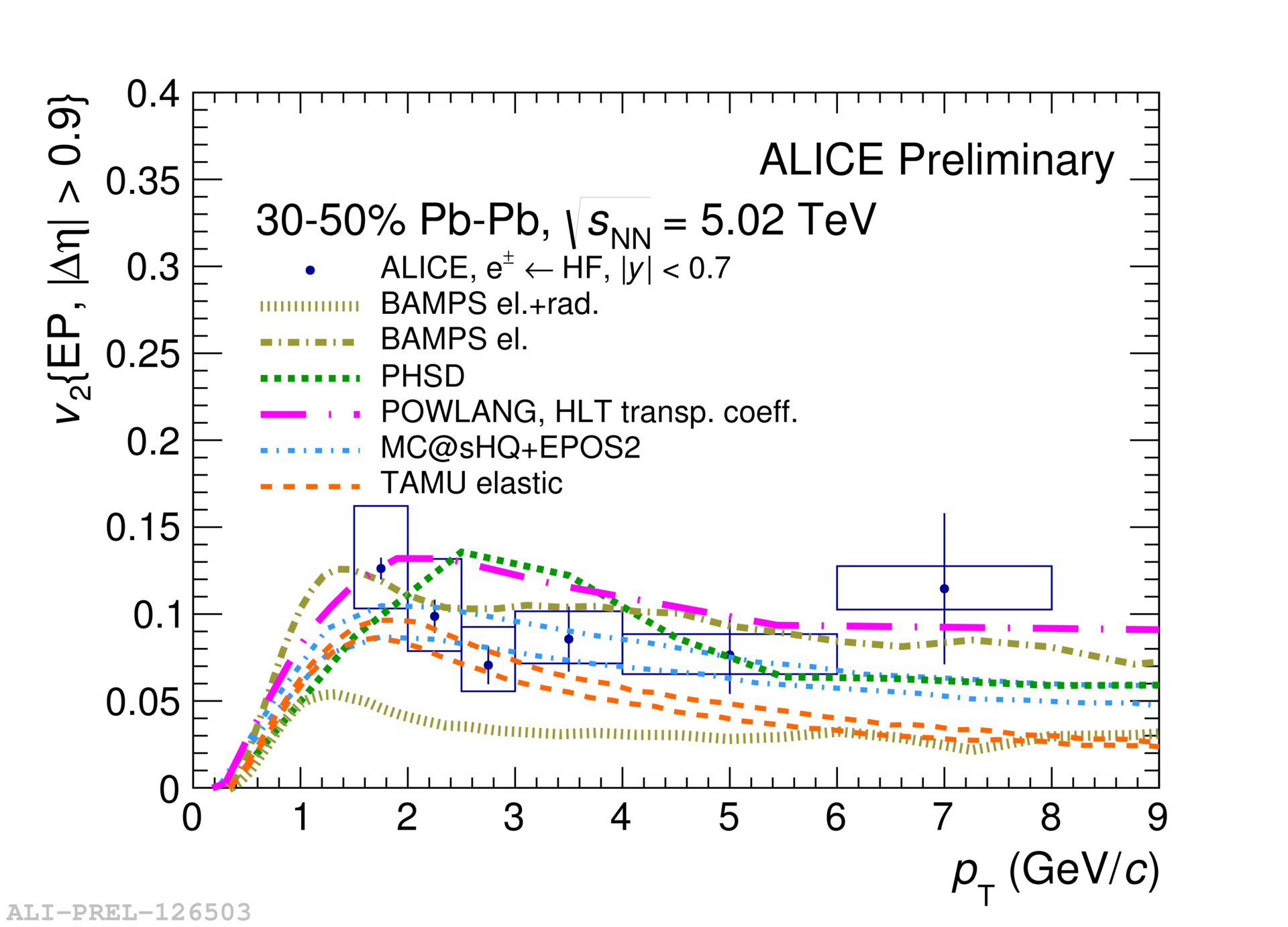}
    \end{subfigure}
    \caption{Left: \raa of electrons from heavy-flavour hadron decays as a function of \pT  in 30--50\% Pb--Pb collisions at \snn = 5.02 TeV. Right: \vtwo of electrons from heavy-flavour hadron decays as a function of \pT  in 30--50\% Pb--Pb collisions at \snn = 5.02 TeV.  Results are compared with model calculations \cite{Beraudo2015,PhysRevC.89.014905,PhysRevLett.114.112301,He2014445,PhysRevC.92.024918,PhysRevC.93.034906}.}
\label{Fig:RAA_v2_HFE_models}
\end{figure}

Figure \ref{Fig:RAA_v2_HFE_models} shows the \pT dependence of the \raae (left panel) and \vtwoe (right panel) measured in 30--50\%  Pb--Pb collisions at \snn = 5.02 TeV compared with model calculations \cite{Beraudo2015,PhysRevC.89.014905,PhysRevLett.114.112301,He2014445,PhysRevC.92.024918,PhysRevC.93.034906}. The models  can describe qualitatively the measurements, although the simultaneous description of the \raae and \vtwoe  measurements remains a challenge for  some of the models.

\section{Conclusions}

The \raa and \vtwo of electrons and muons from heavy-flavour hadron decays have been measured in Pb--Pb collisions at \snn = 2.76 and 5.02 TeV. The \raa results show a strong suppression of leptons from heavy-flavour hadron decays in central Pb--Pb collisions, which is mainly induced by final-state effects due to heavy-quark energy loss in the medium. A positive \vtwo of leptons from heavy-flavour hadron decays is observed in semi-central Pb--Pb collisions, suggesting that heavy quarks participate in the collective motion of the system. The \raa and \vtwo measurements of leptons from heavy-flavour hadron decays show no dependence on rapidity and  collision energies within the uncertainties. 
The presented models can describe qualitatively the suppression and elliptic flow of electrons from heavy-flavour hadron decays, however the simultaneous description of the \raa and \vtwo measurements is a challenge for some of them.

\section{Acknowledgements}

The author was supported by BMBF (FSP201-ALICE).





\bibliographystyle{elsarticle-num}



\end{document}